Spherical-Wave Far-Field Interferometer for Hard X-Ray Phase Contrast Imaging

Houxun Miao[1,*], Andrew A. Gomella[1,*], Katherine J. Harmon[1], Eric E. Bennett[1], Nicholas Chedid[1], Alireza Panna[1], Priya Bhandarkar[2], Han Wen[1,†]

1. Division of Intramural Research, National Heart, Lung and Blood Institute, National Institutes of Health, Bethesda, Maryland, US.

2. Breast Imaging Center, Walter Reed National Medical Center, Bethesda, Maryland, US.

* These authors contributed equally to this work.

† Corresponding author. E-mail: han.wen@nih.gov

Abstract: Low dose, high contrast x-ray imaging is of general interest in medical diagnostic applications. X-ray Mach-Zehnder interferometers using collimated synchrotron beams demonstrate the highest levels of phase contrast under a given exposure dose. However, common x-ray sources emit divergent cone beams. Here, we developed a spherical-wave inline Mach-Zehnder interferometer for phase contrast imaging over an extended area with a broadband and divergent source. The first tabletop system was tested in imaging experiments of a mammographic accreditation phantom and various biological specimens. The noise level of the phase contrast images at a clinical radiation dose corresponded to a 6 nano radian bending of the x-ray wavefront. Un-resolved structures with conventional radiography and near-field interferometer techniques became visible at a fraction of the radiation dose.

The Mach-Zehnder interferometer(*1*) creates wave interference with white light by splitting and recombining a collimated beam along two paths of equal lengths. Coherent interference occurs when the difference in travel times is within the short temporal coherence of the light. It is embodied in modern applications such as optical coherence tomography(*2*), matter wave interference of electrons, neutrons and atoms to study the particle-wave duality and as precision measurement devices(*3-5*), and x-ray phase contrast imaging(*6-8*). Hard x-ray phase contrast techniques have flourished owing to the potential for enriched information at low radiation dose. The x-ray Mach-Zehnder interferometers have the highest levels of phase contrast among a spectrum of techniques(*9, 10*). However, both x-ray and matter wave versions use slits to collimate divergent sources(*3, 4, 10*). This curtails the flux from the source and restricts the area of detection.

A spherical wave from an area source is the sum of a continuous and divergent array of parallel beamlets (Fig. 1). The width of each beamlet equals the transverse coherence of the source at the detector. The spherical-wave Mach-Zehnder (SMZ) interferometer contains a sequence of three phase-shift gratings, with pitches of 400 nm, 200 nm and 400 nm in our implementation. The gratings diffract the beamlets into a cascade of light paths. Among them are pairs of equal lengths capable of coherent interference. Each pair functions as a white-light Mach-Zehnder interferometer, called an MZ pair. Two MZ pairs are illustrated in Fig. 1. Any location on the detector plane is reached by multiple MZ pairs from different beamlets. Although their interference patterns cancel out under perfect symmetry, they can be brought into alignment everywhere on the detector plane by a small adjustment from perfect symmetry. This allows the observation of wave interference in an extended image area. The optimal adjustment parameters are given by theoretical formulas in Supplementary Materials.

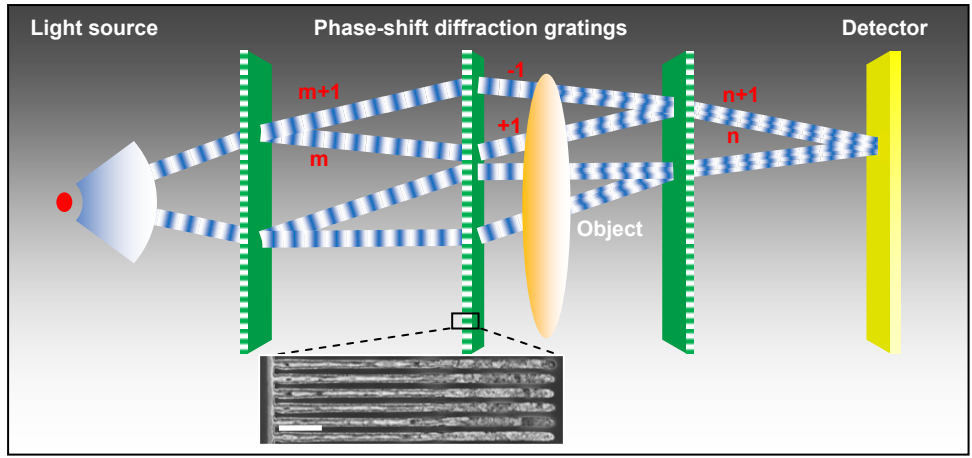

Fig. 1 The spherical-wave Mach-Zehnder interferometer. It has a broadband divergent light source, a series of three $\pi$ phase-shift gratings, and an area detector. The middle grating has a period $P$. The first and third grating have periods of $2(P+\Delta P_1)$ and $2(P+\Delta P_3)$, respectively. The inter-grating distances are $D$ and $(D+\Delta D)$, respectively. An electron micrograph of a cross-section of the mid grating is shown with a scale bar of 600 nm. For any location on the detector, there are multiple pairs of mutually coherent diffraction paths that connect it back to the source. Two such pairs are sketched. One of them consists of the paths with grating diffraction orders of (m+1, -1, n+1) and (m, +1, n). The lateral separations between the paths are exaggerated for the purpose of illustration. All such pairs undergo the +1 and -1 diffraction orders at the middle grating to form a parallelogram between the first and third grating. Each pair functions as a white-light Mach-Zehnder interferometer. In perfect symmetry where all the $\Delta$ values are zero, the interference fringes of all the pairs are out of phase and cancel each other on the detector, resulting in no observable wave interference. However, by adjusting the $\Delta$ values to deviate slightly from perfect symmetry, the interference fringes from the multiple pairs are brought into alignment at all locations on the detector, giving a full field of interference pattern without the need to collimate the light.

A sample placed in the x-ray field causes differential phase shifts and variable loss of mutual coherence between the two light paths of an MZ pair (Fig. 1), resulting in a change of the interference pattern on the detector. The interference fringes are measured in a phase stepping process(*11-13*). The phase differential and the loss of coherence are quantified into a differential phase contrast image and a coherence image (sometimes called the dark field(*14*) or scattering image(*15*)).

The period of the interference fringes on the detector is expressed as $P/(a\Delta_g+b\Delta_D)$, $P$ being the pitch of the middle grating, $\Delta_g$ being the percentage adjustments of the pitches of the first and third gratings, $\Delta_D$ being the percentage mismatch of the inter-grating distances, and $a$ and $b$ are geometric factors related to the distance ratios between the components of the interferometer. The fringe period was 0.27 mm in our hard x-ray implementation.

A condition on the size of the source spot is that the transverse coherence length of the beam at the first grating should not be much less than the pitch of the grating. Thus, the smaller the grating pitches, the larger is the permissible area of the source. The coherence length is approximately $L_1\lambda/S$, $L_1$ being the distance between the source spot and the first grating, and $S$ the size of the source spot. In the current implementation the source spot size was 60 µm in the direction perpendicular to the grating lines.

In one test the SMZ interferometer was applied to imaging the American College of Radiology (ACR) mammographic accreditation phantom model 156. The phantom simulates the compressed breast tissue in mammography screening exams. It is the standard target for quality assurance tests. It has three groups of embedded structures that simulate fibrous tissue, micro calcification and tumors in the breast on a scale from over a millimeter down to 0.16 mm. In a comparison with a modern digital mammography scanner (GE Senographe Essential), the mammography scanner was operated in a standard clinical protocol at an average glandular radiation dose (AGD) of 1.23 mGy and corresponding entrance skin exposure (ESE) of 4.98 mGy. The smallest fibrous feature (0.4 mm diameter) and tumor mass (circular lens shaped mass of 0.25 mm thickness and a few mm diameter) were at the detection threshold, while the smallest calcification feature (0.16 mm) could not be detected (Fig. 2A). This is a high level performance among digital mammography systems(*16*). Using the SMZ interferometer at 78% the AGD dose (0.96 mGy), the smallest features were all resolved with additional details and specific defects of the phantom (Fig. 2B and 2C). The smallest micro calcification and tumor mass were visible at 26% dose (0.32 mGy) (Fig. 2D). The smallest fibrous feature was apparent at 1/12th the dose (0.10 mGy) (Fig. 2E). The micro calcification features were detected in the coherence (dark field) images, indicating that they disrupt the transverse coherence of the x-ray wave. The noise floor of the differential phase images was measured as the standard deviation in featureless areas of the phantom. At the 0.96 mGy AGD dose the noise floor corresponded to a 5.7 nano radian bending of the wavefront. The lower doses increased the noise floor, but the signals from the features remained above the noise floor (Fig. 2F).

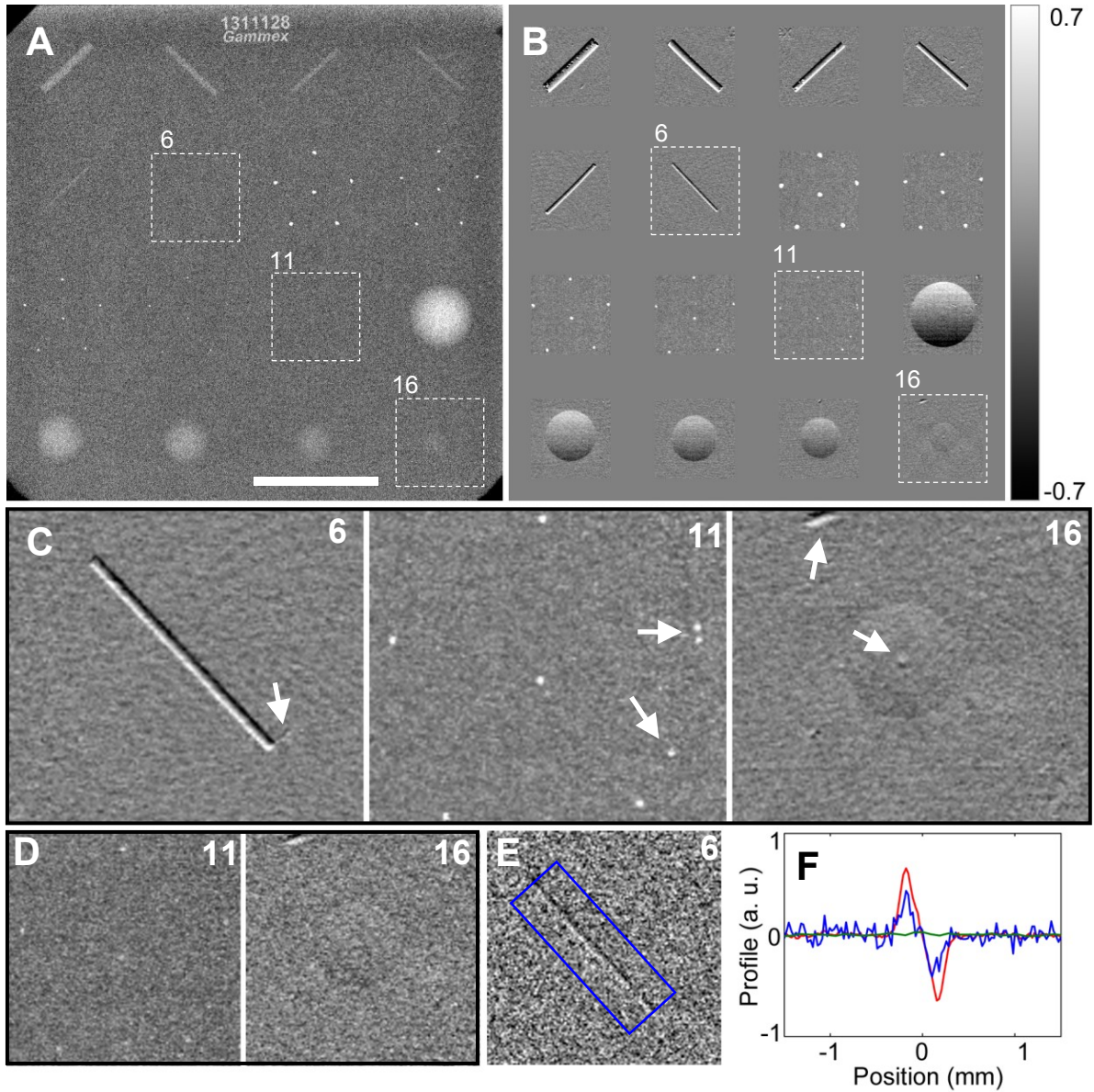

Fig. 2 Comparison of images and radiation doses between the spherical-wave Mach-Zehnder interferometer and a modern digital mammography scanner. (**A**) The digital mammography scanner acquired an attenuation image of a mammographic accreditation phantom using a standard protocol at the average glandular dose (AGD) of 1.23 mGy. The phantom has embedded nylon filaments, aluminum oxide specks and lens-shaped masses that simulate fibrous tissue, micro calcification and tumors, respectively. Features 6, 11 and 16 are the smallest filament (0.4 mm), specks (0.16 mm) and mass (circular lens of 0.25 mm thickness and few mm diameter). Features 6 and 16 are at the detection threshold while 11 is not visible. Scale bar is 2.0 cm. (**B**) A montage of images from the SMZ interferometer at 78% dose (0.96 mGy). The speck features are shown in dark-field (coherence) images. All features are visible. Images of the smallest features are magnified in (**C**), where additional details and defects are indicated by arrows. (**D**) At 26% of the mammography scanner dose (0.32 mGy) the smallest calcification and mass features are visible with the SMZ interferometer. (**E**) The smallest filament feature seen at 1/12th the mammography scanner dose (0.10 mGy). (**F**) Signal profiles across the filament

feature 6 averaged over its length. The red and blue traces are from the SMZ interferometer at 78% and at 1/12th the mammography scanner dose. The green trace is from the mammography scanner.

Figure 3 shows examples of imaging mouse organ specimens. The mouse heart and kidney specimens were first fixed in 10% buffered formalin and then immersed in de-ionized water in a plastic container. At a total skin exposure dose of 1.08 mGy the chambers of the heart and the trunk blood vessels connecting to the atria were all apparent in the phase contrast images (Fig. 3A and 3B). The conventional attenuation image was taken at 2.8 times the dose (3.08 mGy) with a modern flat panel detector (Varian PaxScan 3024M). It did not resolve any of the structures (Fig. 3C). In the corresponding images in the kidney specimen, the renal vasculature were seen in the phase contrast images (Fig. 3D and 3E), and only the attached adipose tissue was visible in the attenuation image.

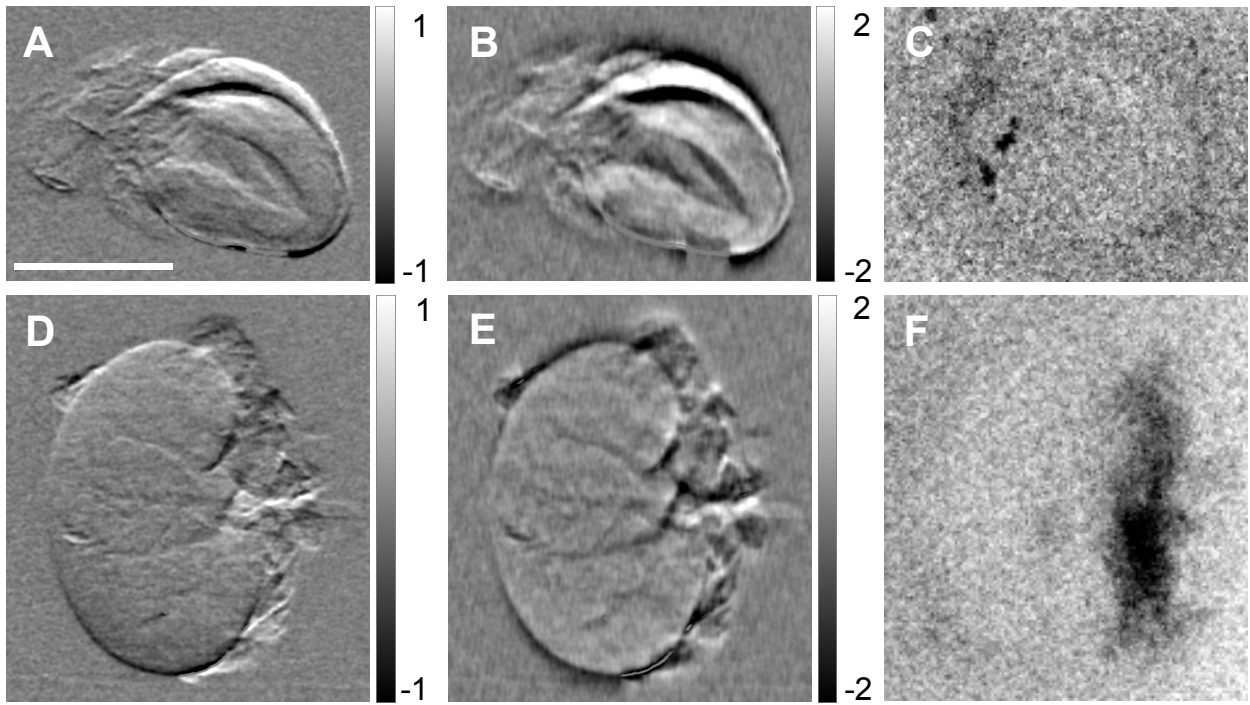

Fig. 3 Comparing phase contrast images of mouse organ specimens from the spherical-wave Mach Zehnder interferometer with attenuation radiography from a modern digital flat panel detector. The phase contrast images were acquired at 35% of the radiation dose of the attenuation radiography images (entrance skin exposure of 1.08 mGy vs. 3.08 mGy). The specimens were fixed in 10% buffered formalin and immersed in water in a 15 mm-thick plastic box. In the differential phase contrast (**A**) and integrated phase shift (**B**) images of a mouse heart specimen, the ventricular chambers and major blood vessels connecting to the atria of the heart are apparent, but not resolved in the attenuation image (**C**). The scale bar is 5.0 mm. (**D**) and (**E**) are the differential phase and integrated phase shift images of a mouse kidney specimen acquired under the same condition. (**F**) is the corresponding conventional radiography at the higher radiation dose. The internal branching structures of the kidney seen in the phase contrast images are renal vasculature. Only the external adipose tissue is seen in the attenuation image.

In a further example of imaging a mouse pup specimen, a one-day old mouse pup was euthanized and then fixed with the same procedure as the organ specimens. The phase contrast images (Fig. 4A and 4B) were taken with the SMZ interferometer at 1.10 mGy skin dose. The conventional attenuation image was taken at 3.08 mGy skin dose (Fig. 4C). Phase contrast showed soft tissue structures which are absent in the attenuation image. The head and abdomen of the mouse are magnified in Fig. 4D and 4E.

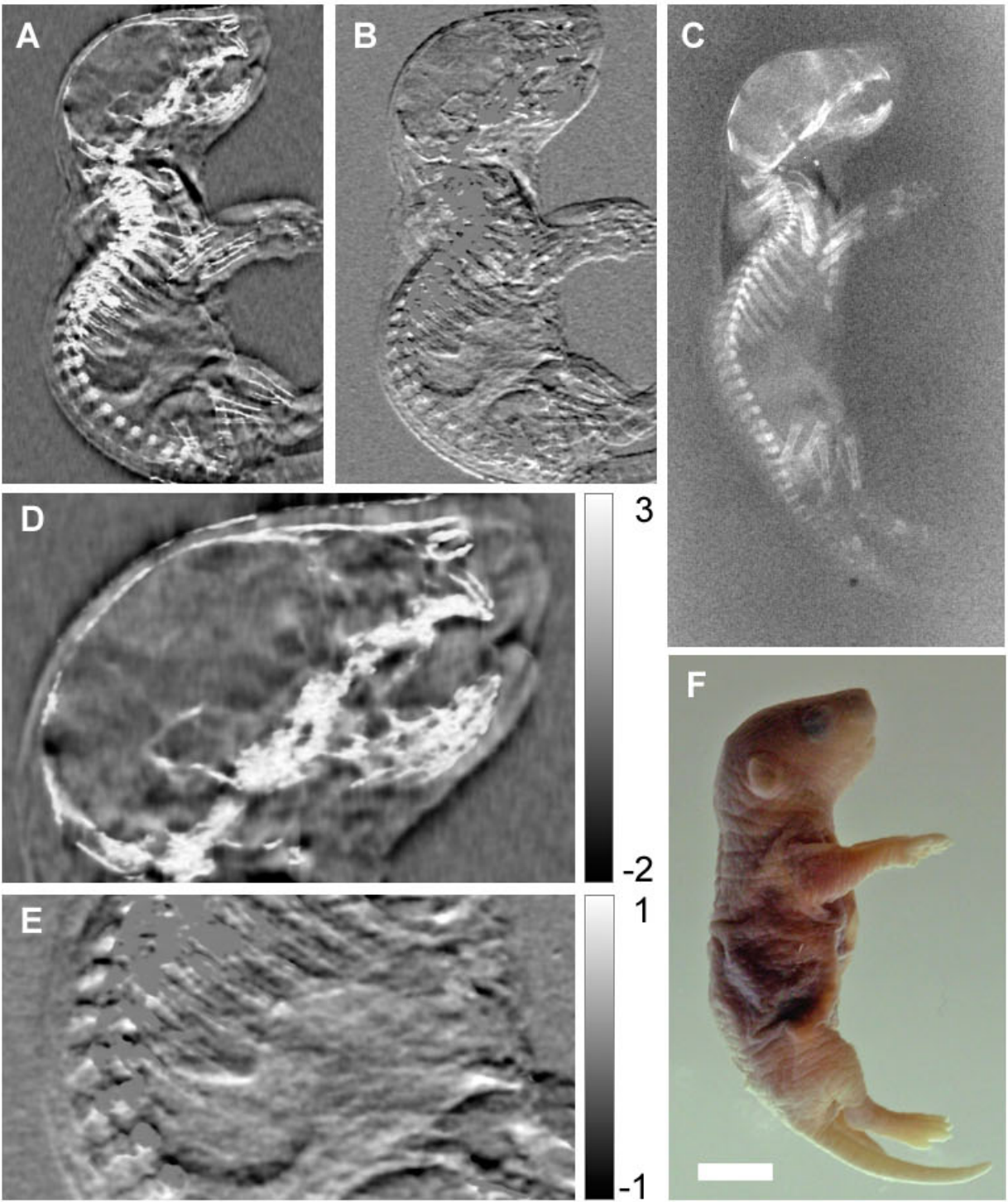

Fig. 4. Phase contrast and conventional radiography of a mouse pup specimen suspended in a water-filled chamber. The spherical Mach-Zehnder interferometer produced the phase shift (A) and differential phase contrast (B) images of a side view at 35% the radiation dose of the attenuation radiography image (C) (1.08 mGy vs. 3.08 mGy entrance skin exposure). Magnified views of the head (D) and abdominal (E) areas of the phase contrast images show soft tissue structures that are not resolved in the attenuation image. (F) A photograph of the specimen chamber with a scale bar of 5.0 mm.

Previously, the near field Talbot-Lau interferometer has been successfully developed for polychromatic and divergent sources in both particle wave and x-ray applications(*17, 18*). However, the x-ray versions used higher than clinical level radiation doses to attain phase contrast information. We compared a Talbot-Lau interferometer with the SMZ interferometer. Both devices are described in Supplementary Materials. The imaged target was the mammographic accreditation phantom. At the same radiation dose of 0.96 mGy AGD, the smallest filament, specks and mass features could not be detected by the Talbot-Lau interferometer. While the noise floors of both devices were similar, there was a 16 fold difference of the phase contrast between the Talbot-Lau and the SMZ interferometers (data presented in Supplementary Materials). The reasons are that the wavelength dependence of the Talbot self-imaging phenomenon limits the sensitivity of the Talbot-Lau interferometer to the near field, and it also requires an absorption grating or mask to readout the Talbot image of another grating, which reduces the photon flux that reaches the detector. In contrast, the Mach-Zehnder interferometer is inherently a broadband far field technique, which allows it to achieve higher levels of phase contrast. It also uses only phase-shift gratings. This reduces the beam attenuation by the gratings, and eases the practical difficulties in making hard x-ray gratings of fine pitches.

In matter wave interferometers, a phase-shift grating can be a laser light wave that coherently diffracts the particle beam(*19-22*). This technique may allow particle interferometers with all laser gratings, to explore greater freedom in controlling the grating parameters and the interaction between the particles and the light field.

In x-ray diagnostic imaging this technique may stimulate other developments besides lower radiation doses. The result from the mammographic phantom suggests that it can be highly sensitive towards microscopic perturbations that collectively disrupt the coherence of the x-ray wave, such as micro-calcification and other heterogeneous changes that occur in tumors(*23*) and vascular plaques, thrombosis and kidney stones. It may therefore help move radiologic imaging to the cellular level, and open possibilities for vessel wall imaging without the use of contrast media. It may also help the development of phase-based contrast agents(*24-28*), and particularly reduce the dose levels of these agents given to patients in potential clinical procedures, such as angiography, targeted imaging of tumors and vascular plaques. Additionally, x-ray mesoscopic imaging commonly uses the cone beam geometry for magnified projections. Here, the spherical-wave Mach-Zehnder interferometer can provide a way to induce strong phase contrast in x-ray projection microscopes.

Acknowledgement

Micro fabrication was performed at the Nanofab Facility of the Center for Nano Science and Technology, National Institute of Standard and Technology. We are grateful to Dr. Leah Zadrozny for preparing the mouse specimens. We are indebted to Mr. Gary Melvin and Dr. Dumitru Mazilu for assistance with mechanical design and machining. We are grateful to Dr. Mark Rivers, Department of Geophysical Sciences and Center for Advanced Radiation Sources, The University of Chicago, for his assistance in writing the instrument control software.

# Supplementary Materials

Materials and Methods

<u>Transition from collimated beam to spherical beam</u>

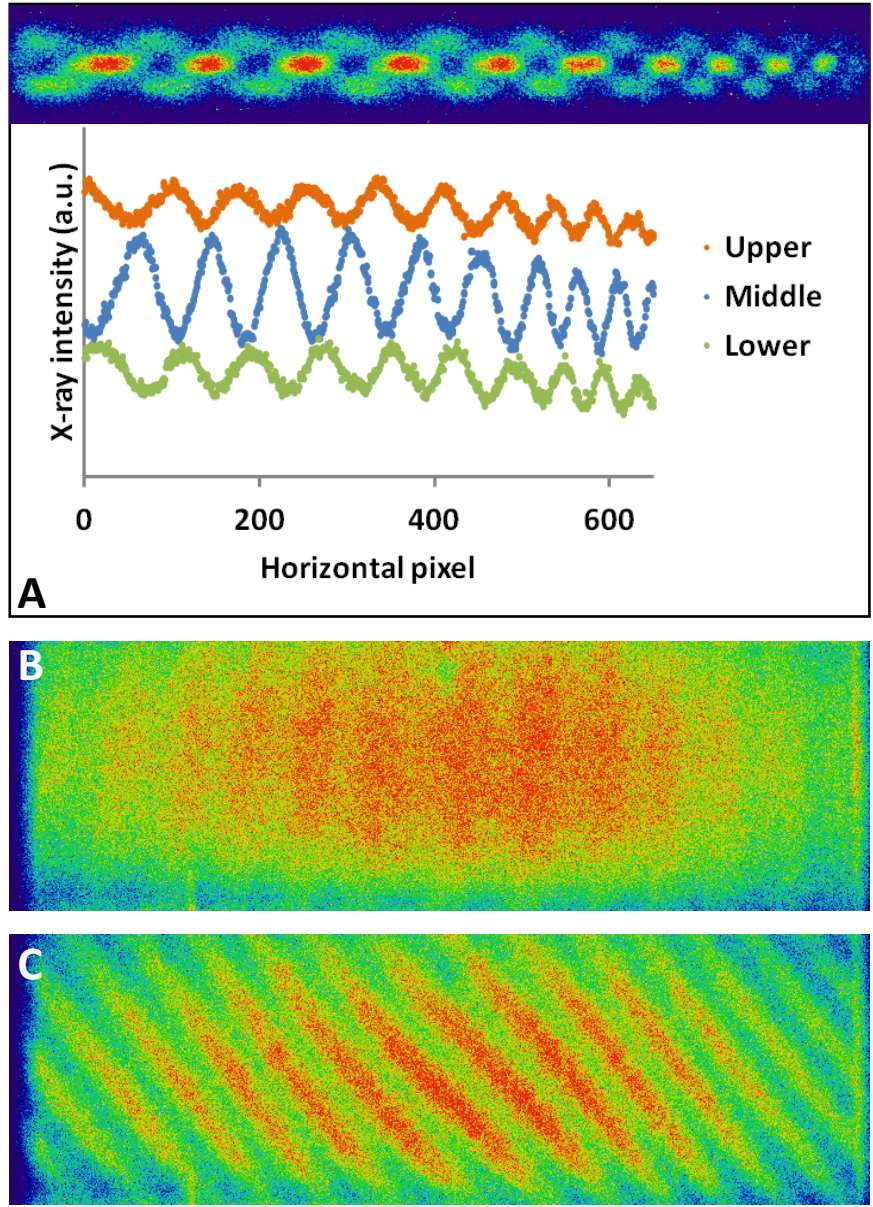

Fig. 1. Transition from a collimated beam to a spherical beam x-ray Mach-Zehnder interferometer. (**A**) Interference fringes of a collimated beam Mach-Zehnder interferometer. Slits are used to collimate the flux of an x-ray tube operating at 50 kVp, and phase shift gratings are used to diffract the collimated beam. The diffraction paths are separated into three horizontal bands on the detector. Each band contains interference fringes from a pair of mutually coherent diffraction paths. The fringes of the middle band are offset by about half a period from the upper and lower bands. (**B**) Upon removing the slits, the bands merge and the interference effect diminishes to a residual level, which comes from the residual intensity modulation of the phase-shift gratings. (**C**) By a small adjustment of the position of the third grating, the fringes of the multiple bands were brought into alignment everywhere on the detector, and the interference effect is recovered in the full field of view.

The transition from a collimated beam to a spherical beam x-ray Mach-Zehnder interferometer is demonstrated in Fig. S1. Using slits to collimate the flux of an x-ray tube, the diffraction paths fall on three separate bands on the detector. Each contains a pair of mutually coherent paths giving rise to interference fringes. Fringes in the brightest central band are shifted by about half a period from those in the side bands. Upon removal of the slits, the bands overlap and cancel each other resulting in the loss of interference effect. This is explained by the theoretical result that in a perfectly symmetrical grating arrangement, the sum of all the MZ pairs relates to the form factor of the intensity transmission functions of the first and third gratings. Since they are phase-shift gratings, there is minimal intensity modulation.

However, noting in Fig. 1 that the different MZ pairs pass through different locations on the gratings to reach the same location on the detector, their relative phases can be altered by modulating the phases of the grating patterns. This is done by a small mismatch of the grating periods and a slight imbalance of the inter-grating spacings. An optimal adjustment is found for all locations on the detector. An example of the recovery of the interference effect is shown in Fig. S1. The optimal adjustment includes a fractional change of the pitches of the first and third gratings by the amount of $f_1 P^2/(2\lambda L_1) + f_3 P^2/(2\lambda L_3)$, $\lambda$ being the central wavelength of the x-ray spectrum, $L_1$ and $L_3$ being the source-to-first grating distance and the detector-to-third grating distance, and $f_1$ and $f_3$ are form factors of the first and third gratings determined by the actual shape of their complex transmission functions; a corresponding mismatch between the inter-grating spacings by $C_1 f_1 P^2/\lambda - C_3 f_3 P^2/\lambda$, $C_1$ and $C_3$ being geometric factors related to the distance ratios between the hardware components. The theoretical modeling that provided the formulas of the optimal adjustment parameters will be presented in a separate paper.

Imaging systems and experimental procedures

The tabletop SMZ interferometer included a tungsten anode x-ray tube operating at 40 kVp or 50 kVp and 1.0 mA current (Sourceray Inc., NY, US), three π phase-shift gratings of 2:1:2 pitch ratios with the central grating pitch of 200 nm, a custom x-ray detector consisting of a GdOS phosphor screen that converted x-rays to green light which was captured by a digital camera with a wide-aperture lens (Nikon D800 SLR, Japan). The x-ray tube had no additional filtering besides its glass wall and a 0.5 mm thick fiber glass layer on the tube window. The focal spot of 0.15 mm exceeded our requirement of 60 μm, which was resolved by placing a 60 μm wide tungsten aperture on the tube window. The detector matrix was 1237 x 820 with a pixel size of 52 μm. The detector was tilted to increase the resolution in the vertical direction perpendicular to the grating lines. The distance between the source and the first grating was 28 cm, the inter-grating distance was 46 cm, and the distance between the third grating and the detector was 53 cm. The total length of the system was 1.73 m. The tuning process for maximal interference effects included adjusting the longitudinal position of the third grating in a range of several millimeters, and tilting the first and third gratings by up to 1.0 degree around a horizontal axis to effectively change their pitches. The fringe visibility, defined as the ratio between the amplitude of the sinusoidal intensity oscillation and the average intensity, was 10% at 40 kVp and 8% at 50 kVp. Each exposure was between 5 sec and 20 sec long, and each phase stepping set included 4 to 12 exposures depending on the radiation dose level. The reported doses are the sum of all exposures. An adaptive image processing method was employed to deal with mechanical drifts in the device(*29*).

The gratings were all rotated around the vertical axis by 60° to double their effective depths for sufficient phase modulation of the x-ray wave. Fabrication of the nanometric hard x-ray phase gratings is described in detail in (*30*). Briefly, electron beam lithography was used to pattern a master template (Eulitha AG, Switzerland) for nano-imprint lithography. Following nano-imprint lithography on a silicon wafer, trenches were etched down to 3.8 µm via cryogenic reactive ion etching. A thin layer of platinum was deposited via atomic layer deposition (ALD), followed by conformal electroplating of gold to fill the trenches. In this first iteration, the grating area was limited by the cost of the electron beam lithography step to 1 mm wide and 7 cm long. Ongoing development of other fabrication techniques for nanometer-scale structures(*31*) may also help to increase the grating depth for higher energy applications.

The digital mammography system used to image the mammographic accreditation phantom was the GE Senopgraphe Essential model. It was operated in a standard clinical protocol, at 29 kVp tube voltage and 56 mAs in a Rh target/Rh filter configuration. Its digital flat panel detector has a pixel size of 100 µm.

The tabletop Talbot-Lau interferometer used to image the mammgraphic accreditation phantom had the same type of x-ray tube as the SMZ interferometer. It had a sequence of three gratings. The source grating and the detector analyzer grating were gold absorption gratings of 4.8 µm pitch and 50 µm nominal depths (Microworks GmbH, Germany). The middle phase grating was a silicon $\pi/2$ phase-shift grating of 2.4 µm pitch. All gratings were rotated by 30° about the vertical axis to increase the effective depth of the absorption gratings. The detector was a digital flat panel detector (Varian PaxScan 3024M) of 83 µm pixel size. The inter-grating distance was set to 39 cm to satisfy the third order Talbot self imaging condition for the photon energy of 22 keV. The fringe visibility was 20% at 40 kVp tube voltage. The total length of the system was 1.02 m. Each exposure was 4 sec long, and a phase stepping set consisted of 19 exposures.

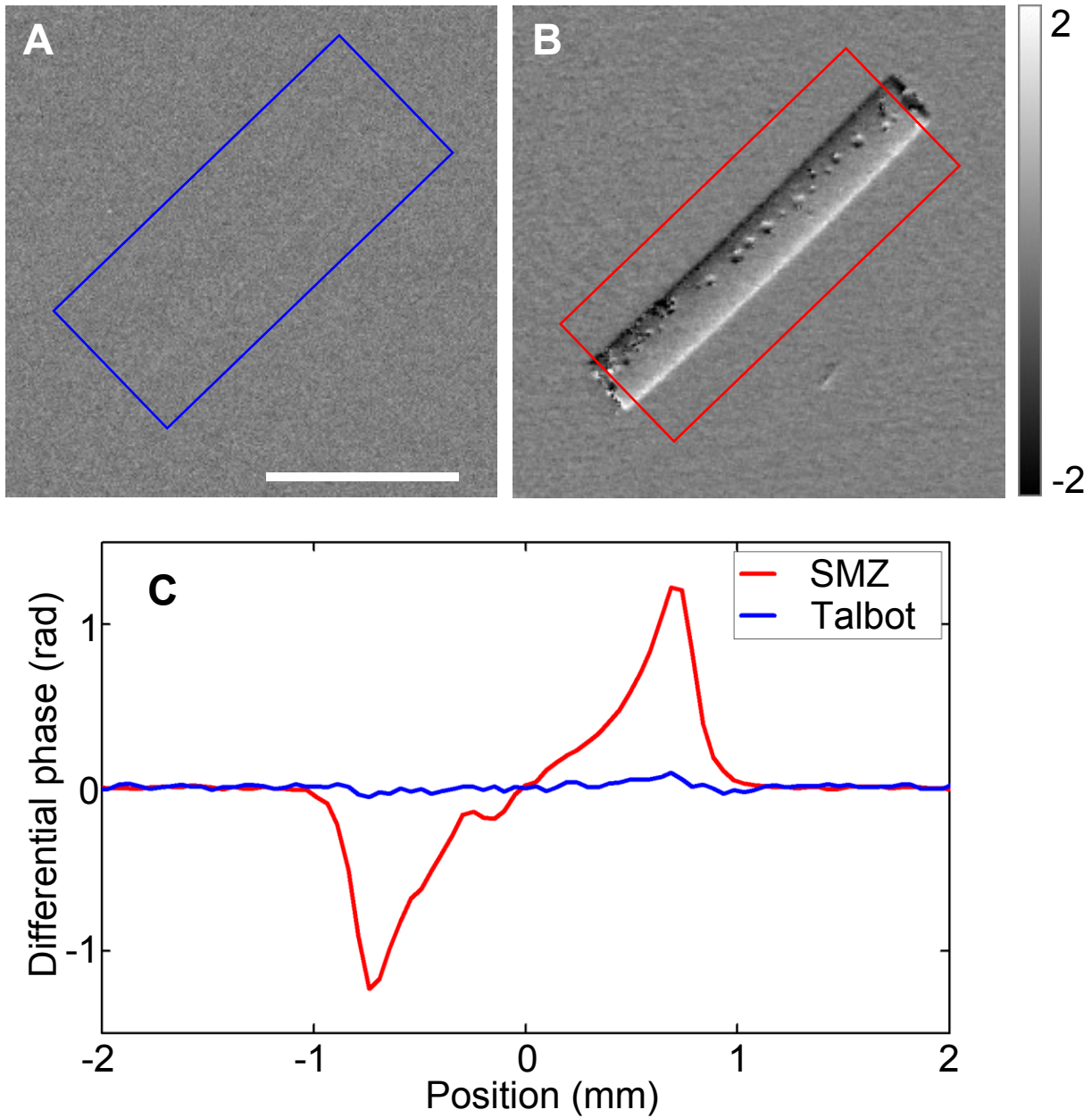

Fig. S2 Comparison of the levels of phase contrast between the Talbot-Lau and Mach-Zehnder interferometers. The differential phase contrast images of the largest filament feature in the mammographic accreditation phantom were taken with (**A**) the Talbot-Lau intereferometer and (**B**) the spherical-wave Mach-Zehnder interferometer. Both had average glandular dose of 0.96 mGy. The scale bar is 5.0 mm. A number of defects in the phantom that overlap the nylon filament are visible in the SMZ image. The phase signal profiles across the width of the filament from both devices are plotted in (**C**).

A comparison between the levels of phase contrast of the spherical Mach-Zehnder and the Talbot-Lau interferometers is shown in Fig. 2S. The mammographic accreditation phantom was imaged with both devices at the x-ray tube voltage of 40 kVp and a radiation dose of 0.96 mGy AGD.

Conventional attenuation radiography images of all samples except the mammographic phantom were acquired with the digital flat panel detector described above. All specimens were placed directly on the surface of the detector to minimize blurring due to scattered radiation.